\def\year{2018}\relax
\documentclass[letterpaper]{article} 
\usepackage{aaai18}  
\usepackage{times}  
\usepackage{helvet}  
\usepackage{courier}  
\usepackage{url}  
\usepackage{graphicx}  
\usepackage{amsmath}
\usepackage{caption}
\usepackage{subfigure}
\usepackage{graphicx}
\usepackage{enumerate}
\usepackage{algorithm}
\usepackage{array}
\usepackage{algorithmic}
\usepackage{multirow}
\usepackage{threeparttable}
\usepackage{color}

\usepackage{mmstyles}

\begin{document}

\title{Probabilistic Ensemble of Collaborative Filters\thanks{This work was done when Min was a research assistant with
Lin.}}

\author{Zhiyu Min\\Alibaba Group\\minzy0716@gmail.com
\And
Dahua Lin\\Department of Information Engineering,\\The Chinese University of Hong Kong\\dhlin@ie.cuhk.edu.hk}

\maketitle


\begin{abstract}
Collaborative filtering is an important technique for recommendation.
Whereas it has been repeatedly shown to be effective in previous work,
its performance remains unsatisfactory in many real-world applications,
especially those where the items or users are highly diverse.
In this paper, we explore an ensemble-based framework to enhance the
capability of a recommender in handling diverse data.
Specifically, we formulate a probabilistic model which integrates
the items, the users, as well as the associations between them into
a generative process.
On top of this formulation, we further derive a progressive algorithm
to construct an ensemble of collaborative filters.
In each iteration, a new filter is derived from re-weighted entries
and incorporated into the ensemble.
It is noteworthy that while the algorithmic procedure of our algorithm
is apparently similar to boosting, it is derived from an essentially
different formulation and thus differs in several key technical aspects.
We tested the proposed method on three large datasets, and observed
substantial improvement over the state of the art, including L$_2$Boost,
an effective method based on boosting.
\end{abstract}


\section{Introduction}
\label{sec:intro}

%
Over the past decades, recommender systems have become an inherent
part of e-commerce and many online sharing services.
It was shown in previous study \cite{rendle2009bpr} that targeted recommendation
is an effective way to influence customers or users and promote business.
The key to a successful recommender system is the \emph{relevance}
of the recommendations.
However, identifying products that are relevant to a user's interest
has never been a trivial task.

%
From a technical perspective, recommendations are usually produced
by either content-based fashion or collaborative filtering. The former relies on
comparing the features of an item and those rated by a given user;
while the latter infers the preferences of a user based on those from other users with similar histories.
In real-world applications, collaborative filtering has shown great effectiveness \cite{hu2008collaborative} and its success has been exemplified on the Netflix Prize \cite{bennett2007netflix}.

%
On the other hand, despite the extensive studies devoted to improving
collaborative filters, the performances of current collaborative filtering
techniques remain unsatisfactory in a number of practical applications.
A key challenge lies in the diversity of the real-world data.
Take the books sold on Amazon for example.
Different customers may purchase a book for different reasons --
some may find the story interesting while others are perhaps just
attracted by the authors.
In collaborative filtering, each item is encoded by a fixed vector,
and so is each user. Such a formulation may
not be able to provide the needed capacity to capture the diverse
ways in which users may evaluate an item.

%


A natural idea to improve the capability of handling diverse data is
to construct an \emph{ensemble}, which can combine multiple predictors
specialized in different kinds of data.
Ensemble methods, as an important family of techniques in machine learning,
have been extensively studied \cite{dietterich2000ensemble,friedman2001greedy},
and shown to be very effective for a number of tasks \cite{friedman2000additive,buhlmann2003boosting}.
However, the use of ensemble methods for recommendation remains relatively
limited. Most of the approaches \cite{wu2016cccf,lee2013local} just work in a vanilla weighted-sum fashion.

%
With these challenges in mind, we develop a new method for constructing
a collaborative filter ensemble. Unlike existing methods for this task \cite{lee2013local}, our method is based on
a probabilistic generative model, in which the latent embeddings for both
items and users are generated from a prior distribution, and the predictions
are then derived from a distribution conditioned on the given embeddings.
With this setting, an ensemble can thus be formalized as a mixture model,
of which each component is a probabilistic predictor.
On top of this formulation, we further derive a \emph{progressive} construction
algorithm. This algorithm begins with a basic collarborative filter.
Then in each iteration, it tries to obtain a new one
that is complementary to the current ensemble by learning from
a re-weighted set. The new predictor will then be incorporated into
the ensemble.
It is worth noting that using this progressive method, one need not
prescribe the size of the ensemble, \ie~the number of component filters.
The algorithm can proceed until the performance saturates
or the ensemble reaches the maximum allowed size.

%
We evaluated the proposed method on three large datasets,
\emph{MovieLens} \cite{harper2016movielens},
\emph{CiteULike} \cite{wang2011collaborative},
\emph{Netflix} \cite{bennett2007netflix},
comparing it with other recommendation methods, especially those
for constructing collaborative filter ensembles.
Experimental results showed that our method consistently outperforms
others on all three datasets.
We also conducted a series of ablation studies to investigate
how different modeling choices influence the overall performance.

%


\section{Related Work}
\label{sec:related}


%
%
Among the various approaches for constructing predictor ensemble,
\emph{Boosting} methods such as AdaBoost and L$_2$Boost
\cite{buhlmann2003boosting,buhlmann2007boosting,friedman2000additive}, show great capability and have been widely used in practice.
AdaBoost works in a stage-wise manner.
At each iteration, misclassified samples are assigned with higher weights
and a new predictor is trained on the re-weighted set.
Note that AdaBoost was developed for binary classification and thus is
not directly applicable for collaborative filtering.
But its underlying idea has inspired a series of other boosting formulations.
L$_2$Boost, on the other hand, aims to fit the raw data distribution,
specifically with L$_2$ Loss. Thus it is very suitable for regression tasks.
%
More recently in \cite{Suh2016lensnmf},
\emph{L-EnsNMF} was proposed
for topic modeling. It is an ensemble approach combining Nonnegative Matrix Factorization and gradient boosting.
Besides, general ensemble theories have also been an active topic in recent years
\cite{lacoste2014agnostic,shaham2016deep,lee2016stochastic}.

%
Along another track, \emph{Matrix co-clustering} methods, which were originally introduced to
explore the blocking structure of a matrix, have also been applied to construct ensembles of collaborative
filters \cite{xue2005scalable,xu2012exploration}, driven by the
intuition that for different subgroups of users and items, the matching
relations between them can be different and may need to be modeled differently.
Recently, Lee~\etal proposed \emph{LLORMA}
\cite{lee2013local,lee2016llorma}, with the assumption that
a rating matrix is low-rank locally (\ie~within sub-matrices). 
%
Based on a related idea,
Beutel~\etal proposed \emph{ACCAMS}
together with its Bayesian counterpart \emph{bACCAMS}~\cite{beutel2015accams}.
ACCAMS is an iterative KMeans-style algorithm, while bACCAMS is a
generative Bayesian non-parametric model.
The most recent progress is \emph{CCCF}~\cite{wu2016cccf}, which adopts a probabilistic method
to model overlapped coclusters, and parallelizable MCMC for scalable inference.
We note that the methods mentioned above were mostly devised for approximating
real-valued matrix rather than collaborative filtering implicit feedback.
Our experiments show that direct application of such methods to recommendation tasks
often results in suboptimal performance.

In addition, \emph{Deep Learning} methods are also attracting more and more attention \cite{wang2015collaborative,zheng2017joint}. However, we will not be discussing them since such methods generally require outer features while ours is focused on the ratings alone.

\section{Formulation}
\label{sec:model}

%
%

Collaborative filtering is an important technique for recommendation \cite{hu2008collaborative,rendle2009bpr},
and has been widely used in practical systems.
%

A widely adopted formulation for collaborative filtering is based on embeddings.
Specifically, given a finite set of users and items, respectively denoted by
$\cU = \{u_1, \ldots, u_m\}$ and $\cV = \{v_1, \ldots, v_n\}$, it associates each
user $u_i$ with an embedded vector $\vu_i \in \Rbb^d$ and
each item $v_j$ an embedded vector $\vv_j \in \Rbb^d$.
Then, we can compute the \emph{matching score} between them
based on the dot product between the corresponding embedded vectors, as
$f_{ij} = \vu_i^T \vv_j$.

The embedding vectors are usually obtained by fitting the matching scores
to the observed ratings on a training set. A popular method for this
is \emph{Weighted Matrix Factorization (WMF)} \cite{hu2008collaborative}. It solves the
embedding vectors by minimizing the following objective function:
\begin{equation} \label{eq:wmf}
    \sum_{i=1}^m \sum_{j=1}^n
    c_{ij} \| r_{ij} - \vu_i^T \vv_j \|^2 +
    \frac{\lambda_u}{2} \sum_{i=1}^m \| \vu_i \|^2 +
    \frac{\lambda_v}{2} \sum_{j=1}^n \| \vv_j \|^2.
\end{equation}
Here, $c_{ij}$ is the confidence coefficient for the observed rating $r_{ij}$,
$\lambda_u$ and $\lambda_v$ are regularization coefficients.
Sometimes in practice, a portion of ratings will be omitted in the training stage,
for which the coefficient $c_{ij}$ will be set to zeros.

Let $\mU = [\vu_1, \ldots, \vu_m]$ denote the \emph{user embedding matrix}
of size $d \times m$,
$\mV = [\vv_1, \ldots, \vv_n]$ the \emph{item embedding matrix}
of size $d \times n$,
$\mR \in \Rbb^{m \times n}$ the (observed) \emph{rating matrix}
with $\mR(i, j) = r_{ij}$, and
$\mC \in \Rbb^{m \times n}$ the \emph{confidence coefficient matrix}
with $\mC(i, j) = c_{ij}$.
Then the objective function in Eq. \eqref{eq:wmf} can be rewritten
into a matrix form as
\begin{equation}
    \| \mC \odot (\mR - \mU^T \mV) \|_F^2 +
    \frac{\lambda_u}{2} \| \mU \|_F^2 +
    \frac{\lambda_v}{2} \| \mV \|_F^2.
\end{equation}
Here, $\odot$ denotes element-wise product,
and $\|\cdot\|_F$ the Frobenius norm. The optimal solutions to
$\mU$ and $\mV$ can be solved using alternate coordinate descent.

\subsection{Probabilistic Formulation}
\label{sub:prob}

As mentioned, the expressive power of a classical collaborative filter
is limited when applied to highly diverse data. A natural idea to mitigate
this problem is to construct an ensemble of filters, each specialized to
a certain kind of data. Towards this goal, we explore a probabilistic
formulation of collaborative filters\cite{mnih2008probabilistic}. As we will see, this formulation
can be readily extended to an ensemble-based framework and will result in
an efficient and elegant learning algorithm.

Specifically, from a probabilistic perspective, a collaborative filter
can be viewed as a conditional distribution as $p(r | \vu, \vv)$, where
the distribution of the rating $r$ depends on the $\vu$ and $\vv$, namely
the latent embeddings of a user and an item.
Based on this notion, we can formulate a probabilistic model that explains
the entire generative process, including the generation of the embeddings
and that of the ratings. The model is described as follows:

\begin{enumerate}
\item For each user $u_i$, draw an embedding $\vu_i \sim \cN(\vzero, \sigma_0^2 \mI)$.
\item For each item $v_j$, draw an embedding $\vv_i \sim \cN(\vzero, \sigma_0^2 \mI)$.
\item For each pair $(i, j)$, draw the rating from the conditional distribution
$r_{ij} \sim p(\cdot | \vu_i, \vv_j)$.
\end{enumerate}
Here, $\sigma_0$ is the prior variance of the embeddings.
Given a training set with partially observed ratings, the problem is to estimate
the embeddings $\mU = [\vu_1, \ldots, \vu_m]$ and $\mV = [\vv_1, \ldots, \vv_n]$.
This can be done via \emph{Maximum-A-Posterior (MAP)} estimation, which is to
maximize the objective function given below with respect to $\mU$ and $\mV$:
\begin{equation} \label{eq:pobj}
    \sum_{(i,j)} c_{ij} \log p(r_{ij} | \vu_i, \vv_j)
    + \sum_{i=1}^m \log p_0(\vu_i)
    + \sum_{j=1}^n \log p_0(\vv_j).
\end{equation}
Here, $c_{ij}$ is the confidence coefficient for the pair $(i,j)$, and
$p_0$ is the probability density function for the embedding prior
$\cN(\vzero, \sigma_0^2 \mI)$. The conditional distribution $p$ is formulated
as a normal distribution centered at $\vu_i^T \vv_j$, as:
\begin{equation}
    r_{ij} | \vu_i, \vv_j \ \sim \
    \cN(\vu_i^T \vv_j, \sigma_r^2).
\end{equation}
It is not difficult to see that under this setting, the learning problem as formulated
in Eq. \eqref{eq:pobj} reduces to a \emph{Weighted Matrix Factorization (WMF)} problem
as in Eq. \eqref{eq:wmf}, with the regularization coefficient given by
$\lambda_u = \lambda_v = \sigma_r^2 / \sigma_0^2$.
%

\subsection{Extension to Ensembles}
\label{sub:ensemble}

From a mathematical standpoint, the probabilistic formulation introduced above
is equivalent to the classical WMF formulation.
Moreover, this new formulation can be readily extended to an ensemble model, via
probabilistic mixture modeling.

Specifically, an ensemble of collaborative filters can be considered as a
probabilistic mixture model, comprising $K$ generative components. In particular, each component has
its own set of embeddings for users and items.
According to this setting, the conditional distribution of the rating
for the pair $(i, j)$ is given by
\begin{equation} \label{eq:mixpred}
    p \left(r \mid i, j, \{\mU^{(k)}, \mV^{(k)}\}_k \right)
    = \sum_{k=1}^K \pi_k \cdot p \left(r \mid \vu_i^{(k)}, \vv_j^{(k)}\right).
\end{equation}
Here, $\vu_i^{(k)}$ and $\vv_j^{(k)}$ are respectively the embeddings
for the $i$-th user and the $j$-th item, with the $k$-th component filter,
$\mU^{(k)} = [\vu_1^{(k)}, \ldots, \vu_m^{(k)}]$ and
$\mV^{(k)} = [\vv_1^{(k)}, \ldots, \vv_n^{(k)}]$.
In addition, this mixture model also comes with a prior distribution $\vpi$
over the $K$ components in the ensemble, and $\pi_k$ is the prior
probability for the $k$-th component.

Generally, like other probabilistic mixture models,
the model above can be estimated from a set of training data using
the \emph{Expectation-Maximization (EM)} algorithm \cite{dempster1977maximum}.
In our context, the EM algorithm alternates between \emph{E-steps} and \emph{M-steps}.
Particularly, the \emph{E-step} computes the \emph{posterior probability}
of the pair $(i, j)$ being generated from the $k$-th component, conditioned on
the embeddings produced by the last iteration, as
\begin{equation}
    q_{ij}^{(k)} \propto \pi_k \cdot p(r_{ij} | \vu_i^{(k)}, \vv_j^{(k)}).
\end{equation}
Note that the probability values $q_{ij}^{(k)}$ are normalized such that
$\sum_{k=1}^K q_{ij}^{(k)} = 1$ for every pair $(i, j)$.
These probability values can also be considered as a set of \emph{``soft weights''}
that associate each pair with different components.

The \emph{M-step}, instead, updates the embeddings for each component
based on a re-weighted WMF. Particularly, for the $k$-th component,
it can be updated by solving the following problem.
\begin{equation}
    \sum_{i=1}^m \sum_{j=1}^n
    c_{ij} q_{ij}^{(k)} \| r_{ij} - \vu_i^T \vv_j \|^2 +
    \frac{\lambda_u}{2} \sum_{i=1}^m \| \vu_i \|^2 +
    \frac{\lambda_v}{2} \sum_{j=1}^n \| \vv_j \|^2.
\end{equation}
Compared to the standard WMF given by Eq. \eqref{eq:wmf}, the only change here is that
the coefficient $c_{ij}$ is replaced by $c_{ij} q_{ij}^{(k)}$.
Therefore, the problem can be solved similarly.
The M-step also updates the prior probabilities as
$\pi_k \propto \sum_{(i,j)} c_{ij} q_{ij}^{(k)}$.


\section{Progressive Construction Algorithm}
\label{sec:method}

While the EM algorithm presented above is a standard way to
estimate mixture models, we found empirically in our study that
it does not work very well for the recommendation task.
A key issue here is that the component filters initialized randomly
(\ie~on a random partition of the ratings) are not sufficiently complementary.
While the E-M updates can optimize them to a certain extent,
they are pretty much trapped in a suboptimal configuration.

For an ensemble model to work, the key is \emph{complementarity},
namely, to maintain a diverse set of components that are \emph{complementary}
to each other.
In this work, we develop a new algorithm for constructing an ensemble of
collaborative filters.
Instead of having all components randomly initialized at the very beginning
and then performing iterative updates,
it constructs the ensemble in a progressive manner, that is, to add one
component at a time. Each new component will be learned to \emph{complement}
the current ensemble, with focus placed on those entries that are not
well predicted.

\subsection{Overall Procedure}


Specifically, this algorithm begins with a standard WMF predictor, with all observed entries weighted equally.
The first component may produce reasonable predictions for a considerable
portion of ratings. However, for a diverse dataset, there can be certain
parts that are not well predicted. New components will be introduced
in following iterations to reinforce them.

%
Next, the key question is how to derive a new component that can complement the
current ensemble. In particular, suppose the current ensemble gives
a conditional distribution $p(r | i, j)$, as in Eq. \eqref{eq:mixpred}.
Now, we want to add one filter $p'(r | i, j)$.
If we assign a weight $\alpha$ to the new component $p'$ and thus
a weight $(1 - \alpha)$ to the existing ensemble, then the combined
distribution would be:
\begin{equation}
    p_\alpha(r | i, j) =
    (1 - \alpha) \cdot p(r | i, j) + \alpha \cdot p'(r | i, j).
    \label{eq:ens}
\end{equation}
This is similar to a finite mixture model, except that this mixture
contains exactly two components, an existing ensemble and a new filter,
and that the first component $p$ is fixed.
Hence, we can estimate $p'$ with a restricted version of the EM algorithm:
\begin{enumerate}
\item Compute the association weights to both components for each observed pair
$(i,j)$. Here, we denote the posterior weight to the new component by $\rho_{ij}$,
and thus the weight to the existing ensemble is $(1 - \rho_{ij})$.
\item Estimate the new component $p'$ based on the re-weighted pairs, where
the coefficient for the pair $(i,j)$ is $c_{ij} \cdot \rho_{ij}$.
\end{enumerate}

\subsection{Computing $\rho_{ij}$}

\begin{figure}
	\scriptsize
    \centering
    \includegraphics[scale=0.5]{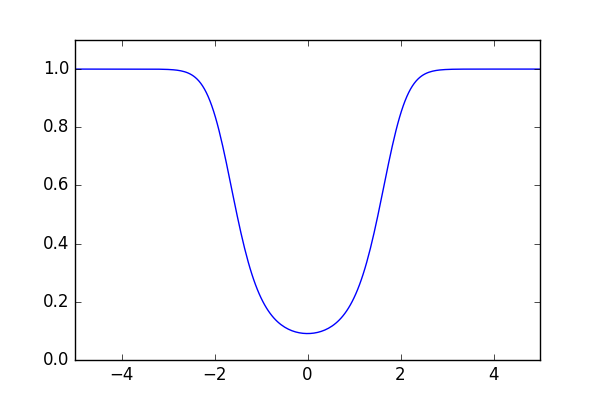}
    \caption{An illustration of $\rho$ as a function of $e$}.
    \label{fig:rho}
\end{figure}

It is important to note that when computing the posterior weights $\rho_{ij}$,
the embeddings of the new component $p'$ are still unavailable and
thus needs to be marginalized out.
The \emph{marginalized} conditional distribution for the new component is
\begin{equation}
    \bar{p}(r | i, j) = \int\int p(r | \vu, \vv) p_0(\vu) p_0(\vv) d\vu d\vv.
\end{equation}
By the symmetry of the formulation, the value of $\bar{p}(r | i, j)$ is a constant
that does not depend on $(i, j)$ and $r$.
Hence, we can simply denote it by $\bar{P}$.
Therefore, with the new embeddings marginalized out, for the pair $(i, j)$,
the posterior probability of being from the new component is given by
\begin{equation}
    \rho_{ij} = \frac{\alpha \bar{P}}{\alpha \bar{P} + (1 - \alpha) p(r | i, j)}.
    \label{eq:reweight}
\end{equation}
Let $\nu = (1 - \alpha) / (\alpha \bar{P})$,
then $\rho_{ij} = 1 / (1 + \nu p(r | i, j))$.
Consequently, the new filter can be constructed based on a re-weighted
distribution, where the ratings at $(i, j)$ is associated with a
coefficient $c_{ij} \rho_{ij}$, as mentioned above.
Note that in our formulation,
$p(r | i, j) \propto \exp(-e_{ij}^2 / \sigma^2)$, where $e_{ij}$ is the prediction error
of the current ensemble for the pair $(i, j)$.
Then, the weight $\rho$ is actually a function of the current prediction error $e$,
as shown in Fig.\ref{fig:rho}.
We can see that when $e$ is relatively small,
$\rho$ increases quadratically as $e$ increases, which means that
the algorithm tends to assign a pair to the new component when it is not
well predicted.
Also, the value of $\rho$ will saturate to $1$ as $e$ continues to increase.

It is noteworthy that the function $\rho$ here is controlled by two
design parameters:
(1) $\nu$ controls the (prior) tendency of assiging samples to the new component;
and
(2) $\sigma$ controls the bandwidth of the weighting curve. With a smaller value
of $\sigma$, the weight $\rho$ would saturate more quickly as the error $e$ increases.


\subsection{Discussions}

Compared to the standard EM algorithm, the progressive method presented above
has several advantages:
\begin{enumerate}
\item It encourages \emph{complementarity} when each new component $p'$ is being
constructed. This is reflected by the weighting function $\rho$. Particularly,
a large prediction error by the current ensemble would result in a greater weight
towards the new component. The experimental results in the next section will show
that this scheme is more effective than EM.

\item It does not require $K$, the number of components, to be specified in advance.
One can continue to add new filters to the ensemble until the prediction performance
is satisfactory or the ensemble reaches the maximum allowed size.

\item The computation complexity for the weights $\rho_{ij}$ is constant, \ie~it does
not depend on the number of components. It makes it scalable to very large ensembles.
\end{enumerate}

The major computational cost of the proposed method lies
at the estimation of new CF components. This cost grows
linearly as the number of components increases. Note that
during inference, all the components can run in parallel, so
the method can scale well in a parallel or distributed environment when deployed for practical service.


\section{Experiments}
\label{sec:experiment}

We conducted experiments to test our method
on three real-world datasets that cover
different application domains and have different rating densities.
In this section, we first introduce the datasets, evaluation metrics,
and the baseline methods,
then present the comparative results and the findings
from ablation studies.
\subsection{Datasets}

\begin{table}[t]
	\centering
	\caption{Data Statistics}
	\label{tab:datastat}
	\begin{tabular}{c|c|c|c|c}
		\hline
		& user\# & item\# & entry\# & density\\
		\hline
		\hline
		MovieLens & 27.7K & 6.4K & 1.8M & 1.00\%\\
		CiteULike & 5.6K & 17.0K & 0.21M & 0.22\%\\
		Netflix & 19.5K & 13.1K & 6.3M & 2.48\%\\
		\hline
	\end{tabular}
\end{table}

\begin{table*}
	\scriptsize
	\centering
	\caption{Recall on MovieLens, CiteULike, Netflix}
	\label{tab:recall}
	\begin{tabular}{c|c|c|c|c|c|c|c|c|c}
		\hline
		\multirow{2}{*}{} & \multicolumn{3}{c}{MovieLens} & \multicolumn{3}{|c|}{CiteULike} & \multicolumn{3}{c}{Netflix}\\
		\cline{2-10}
		& Recall@50 & Recall@100 & Recall@200 & Recall@50 & Recall@100 & Recall@200 & Recall@50 & Recall@100 & Recall@200 \\
		\hline
		\hline
		bACCAMS & 0.213 & 0.348 & 0.525 & 0.073 & 0.125 & 0.220 & 0.080 & 0.151 & 0.270 \\
		\hline
		WMF & 0.377 & 0.540 & 0.698 & 0.388 & 0.494 & 0.594 & 0.142 & 0.271 & 0.452 \\
		\hline
		L$_2$Boost & 0.418 & 0.566 & 0.705 & 0.416 & 0.513 & 0.607 & 0.178 & 0.312 & 0.480 \\
		\hline
		PECF & \textbf{0.460} & \textbf{0.606} & \textbf{0.747} & \textbf{0.446} & \textbf{0.545} & \textbf{0.645} & \textbf{0.258} & \textbf{0.390} & \textbf{0.550} \\
		\hline
	\end{tabular}
\end{table*}

The three datasets in our experiments are described below.

\begin{enumerate}
	\item \textbf{MovieLens}
	is derived from the \emph{MovieLens 20M Dataset} \cite{harper2016movielens}.
	It is a large public benchmark with about $20$ million ratings.
	Removing the users and movies which occur very rarely in the dataset,
	the resultant set contains $27,699$ users and $6,412$ movies.
	We treat all five-star ratings as positive while others as zeros.
	Then the overall rating density is about $1.00\%$.

	\item \textbf{CiteULike}, provided in \cite{wang2011collaborative},
	is a dataset about researchers and the papers they are interested in.
	Researchers can add the articles that they found relevant into their
	respective personal libraries. The author-paper associations in this
	context are typical implicit feedback.
	This dataset contains $5,551$ researchers and $16,980$ papers.
	The overall density is around $0.22\%$.

	\item \textbf{Netflix} is the official dataset
	used in the Netflix Prize competition \cite{bennett2007netflix}.
	The original dataset consists of about $100$ million movie ratings,
	from $480,000$ users and for $17,000$ movies.
	Given the limited computation budget we have, we construct a subset
	with only the frequent users and movies.
	This subset contains $19,455$ users and $13,135$ items.
	Similar to above, we treat all five-star ratings as positive.
	The rating density for this set is around $2.48\%$.
\end{enumerate}

The basic statistics of these datasets are summarized in
Table~\ref{tab:datastat}.
To construct the training, validation, and testing sets,
we randomly split the ratings (both the positive and zeros) into three
disjoint parts by the ratio 3:1:1. These parts are respectively for
training, validation and testing.
We employ NVIDIA Titan X GPUs to accelerate matrix computation.

\subsection{Performance Metrics}

In previous work, various metrics have been used to evaluate the performance of
recommenders. However, as pointed out in \cite{wang2011collaborative},
\textit{recall} is more suitable than \textit{precision} in the context
of recommendation with implicit feedback, as some positive pairs would be
missing from the groundtruths. For example, a user might be truly interested
in an item but did not provide a rating as he/she was unaware of it.
Hence, we follow the metric \textit{recall} in our experiments.
Specifically, for each user,
we sort all the items in descending order of the predicted scores,
and select the top $M$ items to recommend.
\emph{Recall@M} is defined as:
\begin{equation*}
Recall@M=\frac{\textup{number of positive items in top }M}
{\textup{number of all positive items}}.
\end{equation*}
We report the \emph{Recall@M} metric averaged over all users.
In our experiments, $M$ ranges from $50$ to $200$.
It is worth noting that the \emph{Recall@M} metric also imitates
how recommenders are used in the real-world services.

To provide a complementary perspective, we also use
\textit{Weighted Mean Square Error (WMSE)} as an additional metric,
which is defined as
$$
\textit{WMSE} = \frac{\sum_{(i,j) \in \cT} c_{ij}(\hat{r}_{ij}-r_{ij})^2}{\sum_{(i,j) \in \cT} c_{ij}},
$$
where $c_{ij}$ is exactly the confidence coefficient used in WMF.
$(i,j) \in \cT$ indicates that WMSE is evaluated only over the entries in the testing set.
This metric directly measures how well the predicted scores match the ground-truths.

\subsection{Methods to Compare}

Below are all the methods we compared in our experiments.
\begin{enumerate}
	\item \textbf{WMF}.
	\emph{Weighted Matrix Factorization (WMF)} \cite{hu2008collaborative}
	is a popular collaborative filtering technique and often used as a baseline in literatures.
	It derives the latent embeddings for both users and items through regularized
	matrix factorization of the rating matrix $\mR$.

	\item \textbf{ACCAMS}.
	\emph{Additive Co-Clustering to Approximate Matrices Succinctly (ACCAMS)} was proposed
	in \cite{beutel2015accams} for matrix approximation. In this paper, a Bayesian version
	called \emph{bACCAMS} was also developed.
	This method attempts to find local co-clusters of rows and columns to approximate a given
	matrix. Here, we apply to the rating matrix to obtain local collaborative filters.
	In our experiment, we will test the performance of \emph{bACCAMS}.

	\item \textbf{L}$_2$\textbf{Boost+WMF}.
	\emph{L$_2$Boost} \cite{buhlmann2007boosting} is a variant of boosting that aims to minimize the L$_2$ loss.
	It works in a stage-wise manner. In each stage, it fits a new component to the residue of the current
	ensemble. In order to reduce the risk of overfitting and also leave space for subsequent training,
	a shrinkage parameter is usually added for a newly trained predictor.

	\item \textbf{PECF},
	\emph{Probabilistic Ensemble of Collaborative Filters (PECF)} is our method presented in this paper.
	\emph{PECF} constructs an ensemble of filters progressively. In each iteration, a new filter is trained
	on re-weighted entries in order to complement the current ensemble. In the experiments, we want to compare
	it with the methods listed above.
\end{enumerate}


\begin{figure*}
	\centering
	\subfigure[Recall@50 on MovieLens]{
		\centering
		\includegraphics[height=4cm, width=5cm]{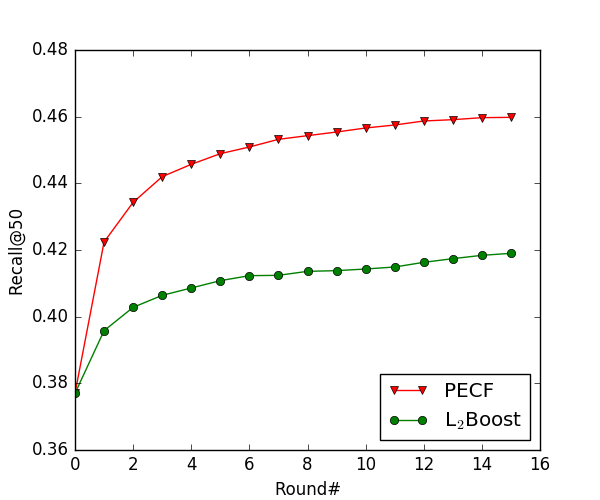}
	}
	\subfigure[Recall@50 on CiteULike]{
		\centering
		\includegraphics[height=4cm, width=5cm]{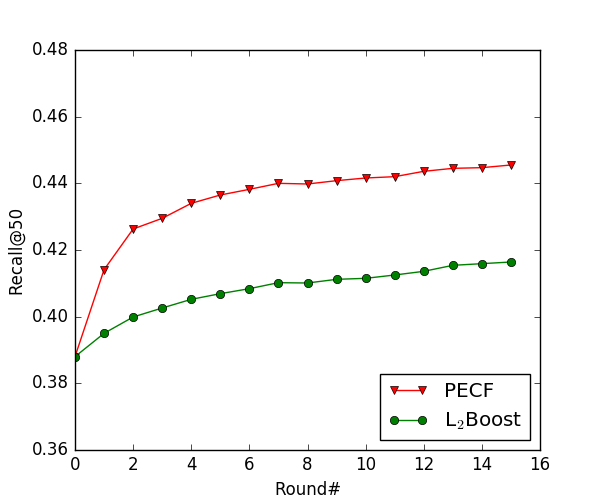}
	}
	\subfigure[Recall@50 on Netflix]{
		\centering
		\includegraphics[height=4cm, width=5cm]{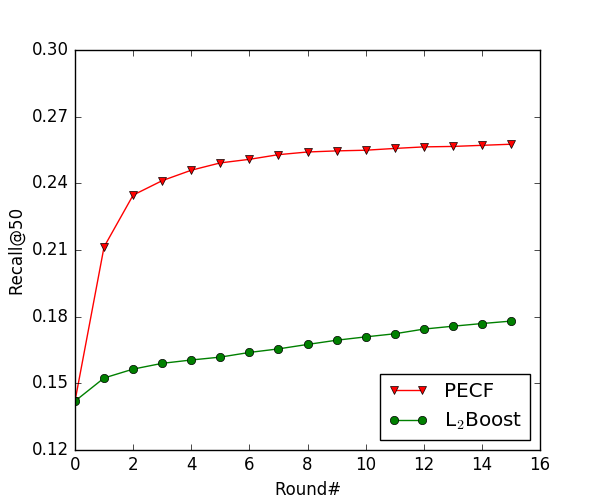}
	}
	\\
	\subfigure[WMSE on MovieLens]{
		\centering
		\includegraphics[height=4cm, width=5cm]{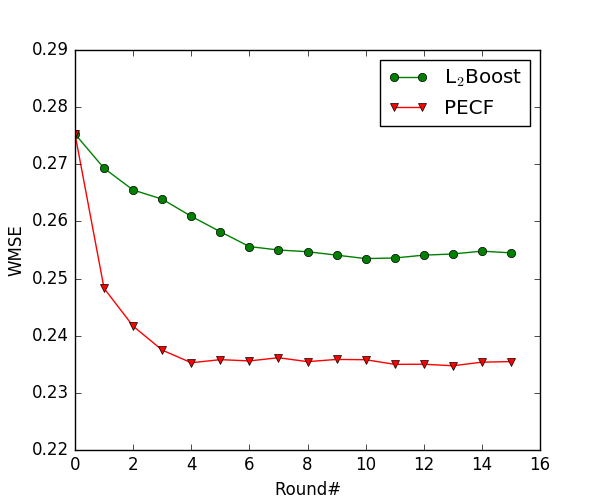}
	}
	\subfigure[WMSE on CiteULike]{
		\centering
		\includegraphics[height=4cm, width=5cm]{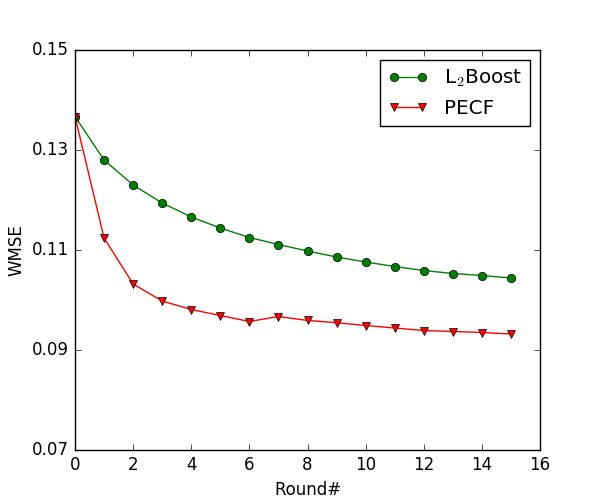}
	}
	\subfigure[WMSE on Netflix]{
		\centering
		\includegraphics[height=4cm, width=5cm]{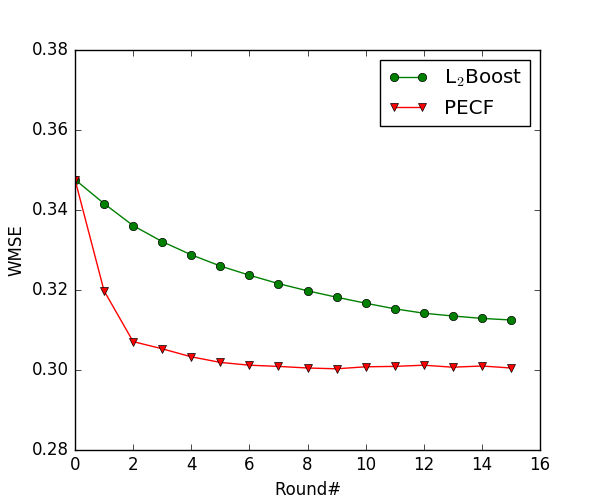}
	}
	\caption{Details in Training Rounds}
	\label{fig:detail}
\end{figure*}

\emph{Detailed Settings:}
In our experiments, we follow the conventional practice to set the confidence coefficient $c$ in WMF,
with $c=1.0$ for positive entries and $c=0.01$ for zero entries.
For bACCAMS, we randomly sample $1\%$ of the missing entries as the negative label.
The latent embedding dimension $d$ is determined respectively on different datasets via cross validation.
Particularly, we set $d$ to $50$ for MovieLens, $150$ for CiteULike, and $50$ for Netflix.
These settings are used for WMF, L$_2$Boost and PECF for fair comparison.
In addition for iterative methods, including L$_2$Boost and our method,
we limit the number of training cycles to be at most $15$.
As for the design parameters $\nu$ and $\sigma$ in our method,
we set them to $\nu=10.0$ and $\sigma=1.0$ via cross validation.

\subsection{Quantitative Comparison}

\begin{table}
	\scriptsize
	\centering
	\caption{Influence of Latent Dimension $d$ on MovieLens}
	\label{tab:Inf_k}
	\begin{tabular}{c|c|c|c|c}
		\hline
		\multicolumn{2}{c|}{ } & Recall@50 & Recall@100 & Recall@200\\
		\hline
		\hline
		\multirow{3}{*}{PECF}& $d=30$ & 0.453 & 0.601 & 0.745 \\
		\cline{2-5}
		& $d=50$ & 0.460 & 0.606 & 0.747 \\
		\cline{2-5}
		& $d=80$ & 0.457 & 0.602 & 0.743 \\
		\hline
		\hline
		\multirow{3}{*}{L$_2$Boost}& $d=30$ & 0.418 & 0.561 & 0.701 \\
		\cline{2-5}
		& $d=50$ & 0.418 & 0.566 & 0.705 \\
		\cline{2-5}
		& $d=80$ & 0.418 & 0.552 & 0.690 \\
		\hline
	\end{tabular}
\end{table}

Table~\ref{tab:recall} compares the performances of different methods on all three datasets
in terms of the \emph{Recall@M} metric. The experimental results suggest:

\begin{enumerate}
\item Ensemble methods, including L$_2$Boost and our method PECF, can notably improve the
recall as compared to the WMF baseline. Particularly, PECF can deliver remarkable performance gains
over a single WMF filter.
On MovieLens dataset, \textit{Recall@50} is promoted from $0.377$ to $0.460$ (around $22\%$ of relative gain).
On CiteULike, \textit{Recall@50} is promoted from $0.388$ to $0.446$ (around $14.9\%$ of relative gain).
The improvement on Netflix is even more significant, where
\textit{Recall@50} is promoted from $0.142$ to $0.258$, with the relative gain at $81.6\%$.

\item Among the three datasets, Netflix has the highest density, but the lowest \textit{Recall}.
It is partly due to the complex and ambiguous patterns in this dataset.
However, our method shows great cability of handling such complicated cases.

\item The performance of bACCAMS, which was originally developed for real-valued matrix approximation,
is clearly inferior to other methods under the
implicit feedback setting. Especially on CiteULike, the extremely low density of the dataset leads to the
poor \textit{Recall} of bACCAMS.
\end{enumerate}

In what follows, we will examine our method in detail, in order to study how different design parameters
influence the overall performance.

\begin{table}
	\scriptsize
	\centering
	\caption{Influence of $\nu$ and $\sigma$ on MovieLens}
	\label{tab:nusigma}
	\begin{tabular}{c|c|c|c}
		\hline
		& Recall@50 & Recall@100 & Recall@200\\
		\hline
		\hline
		$\nu=10,\ \sigma=1$& 0.460 & 0.606 & 0.747 \\
		\hline
		$\nu=10,\ \sigma=0.1$& 0.450 & 0.598 & 0.742 \\
		\hline
		$\nu=1,\ \sigma=1$& 0.435 & 0.590 & 0.738 \\
		\hline
		$\nu=1,\ \sigma=0.1$& 0.431 & 0.586 & 0.735 \\
		\hline
	\end{tabular}
\end{table}

\subsection{Ablation Studies}

We conducted ablation studies to investigate the influence of several design parameters:
the number of components, the latent dimension, the re-weighting parameters $\nu$ and $\sigma$,
as well as the weight of new component $\alpha$.

\begin{figure}
	\centering
	\includegraphics[height=5cm, width=7cm]{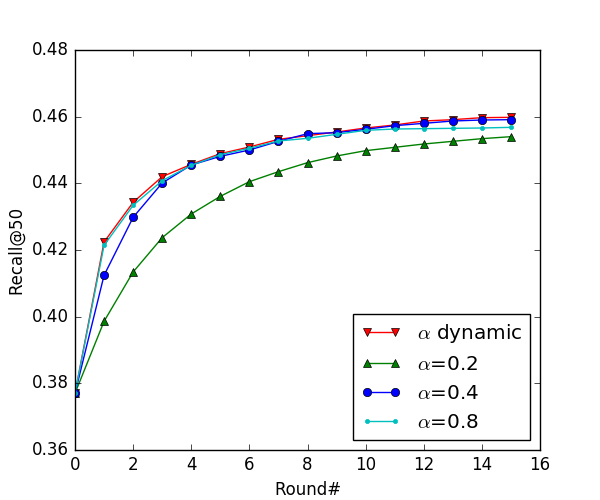}
	\caption{Influence of $\alpha$ on MovieLens}
	\label{fig:alpha}
\end{figure}

\subsubsection{\bf Number of components}

Both L$_2$Boost and PECF construct an ensemble progressively.
They add a new component at each round. Here, we study how the number of new components,
\ie~the number of training rounds, affects the overall performance,
in both \textit{Recall@50} and \textit{WMSE}.
The result is shown in Fig. \ref{fig:detail}.
Note that in these figures, when $\text{Round\#}=0$, the model is exactly a single WMF predictor.

First, we can see that \textit{WMSE} metric is consistent with \textit{Recall@50}.
As mentioned, implicit feedback problems are difficult to tackle.
A common solution is to treat it as a regression task and employ weighted L$_2$ Loss during training.
This consistency between \textit{WMSE} and \textit{Recall} in our experiments
implies that the choice of weighted L$_2$ loss as the training objective is appropriate.

We can also see that the progressive construction leads to stable improvements, both for L$_2$Boost and PECF.
Especially at the first few rounds, the improvements resulted from PECF is substantial.
Compared to L$_2$Boost, PECF not only converges faster, but also outperforms consistently and by a notable margin.

\subsubsection{\bf Latent embedding dimension}

Now we will investigate how the latent dimension $d$ influences the overall performance, for both PECF and L$_2$Boost.

On MovieLens dataset, as mentioned, the best result is attained when $d=50$.
Table~\ref{tab:Inf_k} show the results when $d$ is set to $30$, $50$, and $80$ respectively.
We can see that the results are at a comparable level and not of significant difference.
This implies that naively increasing the latent dimension does not lead to better results.
As we have argued, PECF works in a progressive manner,
and newly trained predictor works as an complementary to the current ensemble.
Whereas the parameter size may be similar to a WMF predictor with higher latent dimension, the key difference lies in the way of how individual components are trained.

\subsubsection{\bf Re-weighting parameter $\nu$ and $\sigma$}

In addition, we also study the influence of re-weighting parameter $\nu$
and the bandwidth paramter $\sigma$ on MovieLens in Table~\ref{tab:nusigma}.
Particularly,
$\nu$ controls the tendency of assigning user/item pairs to new components,
while $\sigma$ determines the bandwidth of the weighting function $\rho$.

We can see that the best result is attained at $\nu=10$ and $\sigma=1$.
The re-weighting curve is shown exactly in Fig. \ref{fig:rho}, where $\rho$ gradually saturates after 2.
In fact, we empirically found that most of the predictions lie before saturation, and
the new weight is very similar to a vanilla quadratic function of the error $e_{ij}$.
Those points residing out of the bandwidth can be viewed as outliers and
the saturation suppressed their impact, thus resulting in more reliable training.

\subsubsection{\bf New component weight $\alpha$}

In our model,
the mixture parameter $\alpha$ is determined dynamically,
in a way that is similar to line search in steepest gradient descent.
In this part, we will see the comparison with fixed $\alpha$,
and the results on MovieLens are shown in Fig. \ref{fig:alpha}.

For fix settings, smaller and larger values of $\alpha$ respectively have their own limitations.
In particular,
when $\alpha=0.2$, the result is still slightly increasing and has not completely converged, and this is surely not an ideal situation as it requires longer training time;
when $\alpha=0.4$, the \textit{Recalls} at the last few rounds are similar to dynamic setting,
but performs worse at the first few rounds;
when $\alpha=0.8$, the initial converging rate is competitive, but the final converge value is not as satisfied.
Overall, setting $\alpha$ dynamically at each round is the best strategy.

\subsection{Comparison to EM Algorithm}

\begin{figure}
	\centering
	\includegraphics[height=5cm, width=7cm]{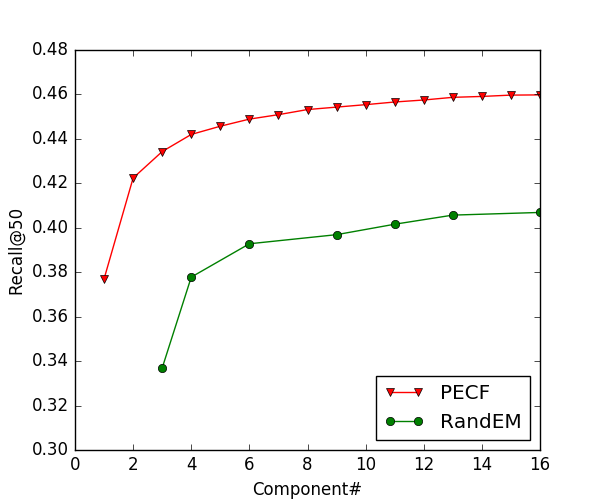}
	\caption{Comparison with RandEM on MovieLens}
	\label{fig:randEM}
\end{figure}

As mentioned, the ensemble is a mixture model that can be trained using the EM algorithm,
first initialized and later updated
in parallel.
The proposed PECF method works in a progressive manner, adding one complementary component at a time.
Fig. \ref{fig:randEM} show the results of both ways on MovieLens,
which compare the performance of both algorithms under
different number of components. In this figure, the EM algorithm with random initialization
is referred to as \emph{RandEM}, and \emph{L$_2$EM} for initialization from L$_2$Boost.

We observe:
(1) Both algorithms can improve the performance as the number of components increases;
(2) RandEM performs substantially worse than our construction method. This is partly ascribed to
the reason that for RandEM, components are randomly initialized, and not explicitly encouraged
to be \emph{complementary}.
This experiment, again, suggests the importance of \emph{complementarity} in ensemble learning.


\section{Conclusion}
\label{sec:conclusion}

This paper presented a new method to construct an ensemble of
collaborative filters for recommondation. It is based on a probabilistic
mixture formulation. But unlike previous work that usually adopts
EM estimation, we develop a progressive algorithm that explicitly
encourages complementarity among component filters.
Experiments on three large datasets, namely MovieLens, CiteULike, and
Netflix, show that the proposed method not only leads to significant
performance gain over a single collaborative filter, but also
outperforms other ensemble-based methods, including L$_2$Boost and bACCAMS,
by considerable margins.
It is noteworthy that our method is a generic approach to ensemble construction.
While we adopt WMF as the basic component in this paper, our method can
also work with other predictors as the basic components.

\section{Acknowledgement}

This work is partially supported by the Big Data Collaboration
Research grant from SenseTime Group (CUHK Agreement
No. TS1610626), the General Research Fund (GRF) of
Hong Kong (No. 14236516).

\bibliographystyle{aaai}
\bibliography{main}

\end{document}